\documentclass[twocolumn,pra, aps,superscriptaddress,showpacs]{revtex4-2}

\usepackage{color}
\usepackage{graphicx}
\usepackage{soul}
\usepackage{amsmath}

\begin{document} 

\title{Coupling light to a nuclear spin gas with a two-photon linewidth of five millihertz} 

\author{Or Katz}
\affiliation{Department of Physics of Complex Systems, Weizmann Institute of Science,
Rehovot 76100, Israel}
\affiliation{Rafael Ltd, IL-31021 Haifa, Israel}
\affiliation{Present address: Department of Electrical and Computer Engineering, Duke University, Durham, NC 27708}
\author{Roy Shaham}
\affiliation{Department of Physics of Complex Systems, Weizmann Institute of Science, Rehovot 76100, Israel}
\affiliation{Rafael Ltd, IL-31021 Haifa, Israel}
\author{Ofer Firstenberg}
\affiliation{Department of Physics of Complex Systems, Weizmann Institute of Science,
Rehovot 76100, Israel}

\begin{abstract}
Nuclear spins of noble gases feature extremely long coherence times but are inaccessible to optical photons. Here we realize a coherent interface between light and noble-gas spins that is mediated by alkali atoms. 
We demonstrate the optical excitation of the noble-gas spins and observe the coherent back-action on the light in the form of high-contrast two-photon spectra. We report on a record two-photon linewidth of $5\pm0.7$ mHz (millihertz) above room-temperature, corresponding to a one-minute coherence time. This experiment provides a demonstration of coherent bi-directional coupling between light and noble-gas spins, rendering their long-lived spin coherence accessible for manipulations in the optical domain.
\end{abstract}

\maketitle

\section*{Introduction}
The coupling of light to atomic spins is a principal tool in quantum information processing using photons \cite{QIP1,QIP2,QIP3,QIP4} and in precision optical spectroscopy, enabling determination of atomic structure \cite{Arimondo,Henderson},  time and frequency standards \cite{J-Ye}, and laboratory searches of
new physics \cite{NP}. 
The performance of these applications depends on the coherence time of the spins and on the efficiency with which they couple to light.
In dense atomic gases, light can couple efficiently to the collective atomic spin of the ensemble \cite{Polzik-RMP-2010}.
However, at room temperature and above, this collective spin is prone to decoherence due to interactions of the atoms with the environment and
due to motional dephasing, which typically limit the coherence time to 10-100 milliseconds \cite{Shaham-2020,relax1,Novikova-review,Walsworth,Firstenberg-RMP}. 
Alkali vapor in anti-relaxation coated cells can reach coherence times as long as one minute \cite{Budker_onemin,Xiao,budker_paraff,Katz_paraffin} and are successfully employed in quantum-optics applications \cite{Polzik-RMP-2010}, but the high-quality coatings degrade at elevated temperatures and thus limit the alkali densities.

\begin{figure}[t]
\centering{}\includegraphics[clip,width=8cm]{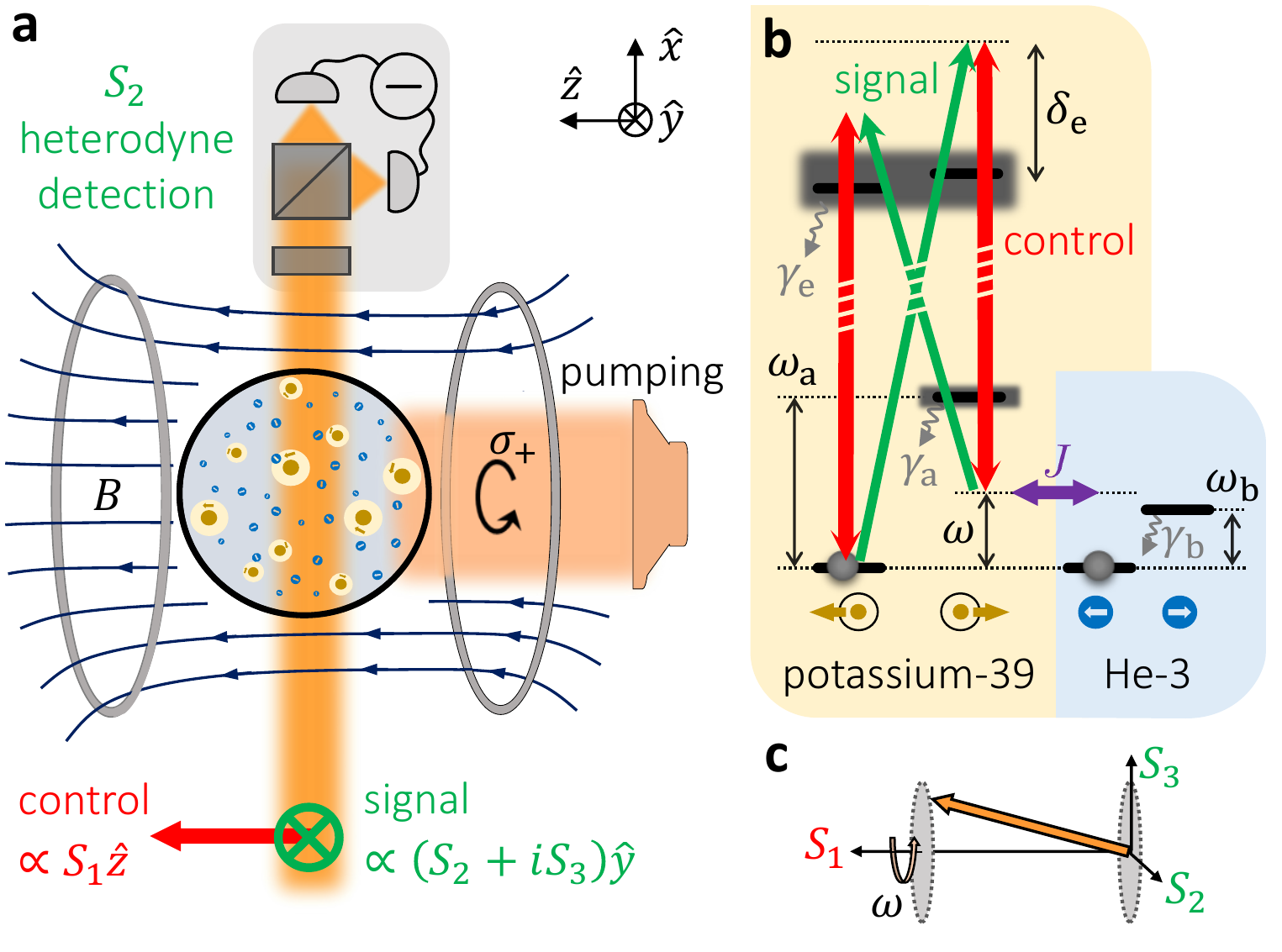}\caption{\textbf{Coupling light to noble-gas nuclear spins.} (\textbf{a}) Experimental setup and (\textbf{b}) level diagram of the optical and spin-exchange couplings. A gaseous mixture of potassium (beige) and helium-3 (blue) atoms is contained in a spherical glass cell. The potassium spins are polarized along the $\hat{z}$ direction by absorption of circularly-polarized pumping light. The helium spins are polarized by collisions with potassium atoms. 
The incoming light comprises a $\hat{z}$-polarized control field (quadrature $S_{1}$ in the Stokes representation) and a $\hat{y}$-polarized signal field (quadratures $S_{2}$ and $S_{3}$). It interacts with the potassium spins via a two-photon process and is measured by heterodyne detection. Both fields are far-detuned by $|\delta_{\mathrm{e}}|\gg\gamma_{\mathrm{e}}$ from the potassium optical transitions of width $\gamma_{\mathrm{e}}$, but the signal field differs from the control field by a small frequency $\omega$, leading to precession in the Stokes vector representation
(\textbf{c}). The alkali and noble-gas spins are coherently coupled by spin-exchange at a rate $J$ and decohere at rates $\gamma_{\mathrm{a}}\gg\gamma_{\mathrm{b}}$. A magnetic field $B\hat{z}$ determines the Larmor frequencies $\omega_{\mathrm{a}}$ and $\omega_{\mathrm{b}}$ of the two spin excitations, thus controlling their mutual coupling and their detunings $\delta_{\mathrm{a}}=\omega-\omega_{\mathrm{a}}$
and $\delta_{\mathrm{b}}=\omega-\omega_{\mathrm{b}}$ from the optical modulation frequency.}
\end{figure}

Odd isotopes of noble gases, such as $^{3}$He, posses
a nonzero spin in their nuclei. The nuclear spin is protected by the
full electronic shells and therefore exhibits extraordinary long coherence
times, possibly many hours. This corresponds to a narrow nuclear magnetic resonance (NMR) that is employed for precision
sensing \cite{romalis-2005,Walker-NMRG}, medical imaging \cite{Walker-RMP-2017},
and searches for new physics \cite{PM2,PM4,PM5,PM6}. As noble gases are transparent to light from infrared to ultraviolet, the preparation and monitoring of their nuclear spins usually rely on collisions with another spin gas \cite{Walker-rmp-1997,meop}. Noble-gas NMR sensors employ spin-exchange collisions with alkali atoms. Since alkali spins do couple to light, the pickup of the NMR signal can be done optically, and in this way narrow spectra and long-lived spin precession signals are routinely obtained \cite{happer-1998,Kornack2002,Walker2015,CHIN}. Yet, various quantum optics applications require an efficient bi-directional coupling between light and noble-gas spins \cite{Firstenberg-QND,SINATRA1,SINATRA2,noble-storage-light,noble-storage-light2}. Such coupling, corresponding to a resonant optical excitation of the long-lived nuclear spins, has never been realized.

Here we realize a coherent bi-direction coupling between
light and noble-gas spins, mediated by alkali spins. We observe a significant attenuation of the optical signal at the nuclear magnetic resonance of the noble gas, accompanied by excitation of noble-gas spin coherence. We control the contrast and
width of the spectroscopic lines with an external magnetic field,
realizing narrow linewidths down to $5\pm0.7~\mathrm{mHz}$ (millihertz). The efficient collective coupling to the noble-gas spins, enabled by the high alkali vapor density, is reflected in the high spectroscopic contrast and in a direct measurement of the transverse spin excitation of the noble gas. This interface between photons and long-lived nuclear spins opens routes to various applications in precision
sensing and quantum information science, including generation of entanglement
and quantum memories with unprecedented lifetimes \cite{Firstenberg-QND,noble-storage-light}.

\section*{Results}
To generate a bi-directional coupling between light and noble-gas spins, we use
a mixture of gaseous $^{3}$He (nuclear spin $R=1/2$) and potassium (electron spin $1/2$ and nuclear spin $3/2$), enclosed
in a spherical glass cell at $187^{\circ}$C, as shown in Fig.~1.
Light directly interacts only with the potassium atoms, whose spin is optically accessible owing to their electronic orbital transitions in the infrared and to their strong spin-orbit and hyperfine couplings. We optically pump the potassium spin $\mathbf{F}$ along $\hat{z}$, which in turn continuously
polarizes the helium nuclear spin $\mathbf{R}$ via spin-exchange collisions
\cite{Walker-rmp-1997}. A constant magnetic field $B\hat{z}$ controls the Larmor frequencies $\omega_{\mathrm{a}}$ and $\omega_{\mathrm{b}}$ of the alkali and noble-gas spins, which are also affected by light shifts and by the equivalent magnetic field produced by the polarized spins.

The light field interacting with the spins is a superposition of a signal field and an auxiliary control field, co-propagating in the
$\hat{x}$ direction. Both fields are tuned near the potassium $D_{1}$ transition at $770~\mathrm{nm}$, which appears as a single optical line of width $2\gamma_{\mathrm{e}}=25~\mathrm{GHz}$ due to pressure broadening by the $^{3}$He at $1500~\mathrm{Torr}$. The fields are detuned by $\delta_{\mathrm{e}}\approx40\gamma_{\mathrm{e}}$ from this line, so their attenuation via scattering is suppressed. The control is linearly polarized parallel to the magnetic field, while the weak signal is linearly polarized in the transverse direction $\hat{y}$. 

The optical frequency of the signal field differs from that of the control field by a tunable frequency $\omega$, generated in our experiment by two acousto-optic modulators. This frequency difference leads to a temporal polarization modulation of the combined field, which can be represented as a precession of the optical Stokes vector, as shown in Fig.~1c. In this representation, the optical signal resides in the oscillating components $S_{2}$ and $S_{3}$. We perform heterodyne detection before and after the cell using two pairs of differential photodetectors to obtain the incoming and outgoing signal components $S_{2}^{\mathrm{in}}(\omega)$ and $S_{2}^{\mathrm{out}}(\omega)$ in a frame rotating at the modulation frequency $\omega$. Technically this is done by extracting the amplitude and phase shift of the measured harmonic oscillations at the frequency $\omega$. This spectroscopic technique, in which light efficiently couples to the orientation moment of the alkali spins irrespectively of the width of their optical transitions, is a special variation of the nonlinear Voigt effect \cite{NMOR} that is applicable even at high buffer-gas pressures. Further details on the experimental configuration and schemes are given in Methods.

We begin our investigation in the regime $\omega_\mathrm{a}\gg\omega_\mathrm{b}$, where the magnetic resonance of the noble-gas spins is detuned away from that of the alkali spins. 
In this regime, the noble-gas resonance is nested in the tail of the alkali resonance, much like a Raman line that is nested in the tail of an optical line. If $|\omega_{\mathrm{b}}-\omega_{\mathrm{a}}|\gg \gamma_{\mathrm{a}}$, where $\gamma_\mathrm{a}$ is the alkali spin decoherence rate, and we tune the optical field near the noble-gas resonance $\omega\approx\omega_{\mathrm{b}}$, then the alkali spins act as an intermediate system that is only weakly excited. Indeed, and despite the large detuning from the alkali resonance, we observe a dramatic change in the signal field due to the nonlinear Voigt effect around the noble-gas resonance. 
We observe a response
in both the transmission $|S_{2}^{\mathrm{out}}(\omega)/S_{2}^{\mathrm{in}}(\omega)|^2$ and phase $\varphi=\mathrm{arg}[S_{2}^{\mathrm{out}}(\omega)/S_{2}^{\mathrm{in}}(\omega)]$ of the $S_{2}$ light quadrature,
as shown in Figs.~2a,b for $B=6.1~\mathrm{mG}$ ($|\omega_\mathrm{b}-\omega_\mathrm{a}| \approx 43\gamma_\mathrm{a}$). We fit the sharp spectral line 
in Fig.~2a to a Lorentzian shape and find a maximal attenuation (contrast)
of $\mathcal{C}=(53\pm3)\%$ and a full width of $2\gamma=10.5\pm1.4~\mathrm{mHz}$. The observed
line center ($\Delta=0$ in Fig.~2) differs from the bare resonance
$\omega=\omega_{\mathrm{b}}$ by ${<}50~\mathrm{mHz}$ predominantly due to NMR shifts induced by the alkali spin-exchange field.
The same parameters $\gamma$ and $\mathcal{C}$ fit as well the phase
shift across the resonance, presented in Fig.~2b, as calculated
from a theoretical model (see Methods).

\begin{figure*}[t]
\begin{centering}
\includegraphics[clip,width=13cm]{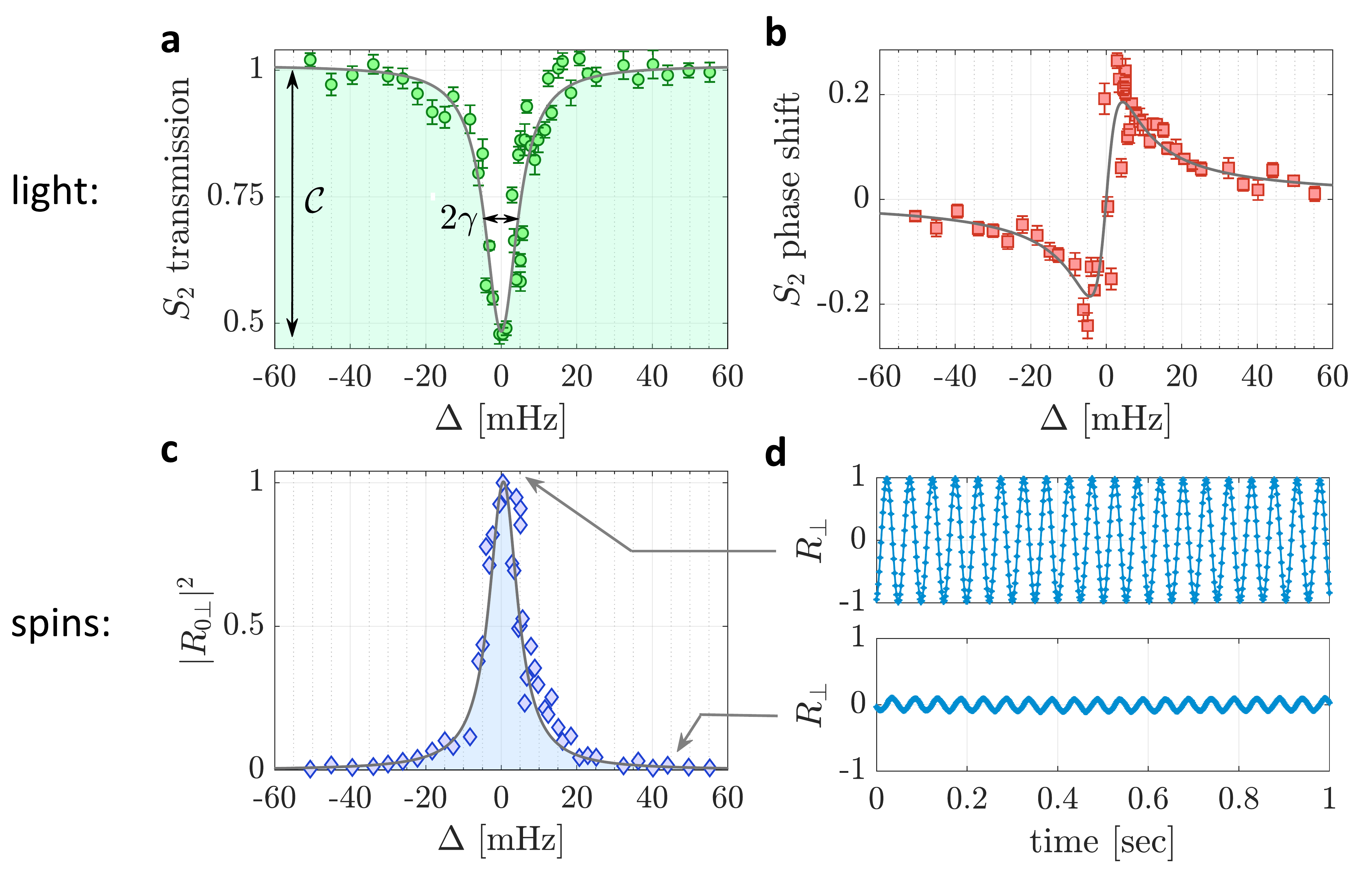}
\par\end{centering}
\centering{}\caption{\textbf{Optical spectroscopy and excitation of helium-3 spins. a}. Transmission of the $S_{2}$ quadrature of the signal light versus the signal frequency
detuning $\Delta$ from the observed resonance. The sharp spectral line,
due to a two-photon process, is obtained when the frequency difference
$\omega$ between the signal and control fields approaches the nuclear magnetic
resonance of the noble gas $\omega_{\mathrm{b}}=19.88~\mathrm{Hz}$
(here $B=6.1~\mathrm{mG}$ and $\omega_{\mathrm{a}}=2.2~\mathrm{kHz}$). A Lorentzian profile (solid line) fits
the data well, providing a narrow linewidth of $2\gamma=10.5\pm1.4~\mathrm{mHz}$
(millihertz) and a contrast of $\mathcal{C}=53\%\pm3\%$. \textbf{b}.
Phase shift of the $S_{2}$ quadrature of the signal. The theory {[}solid
line, see Eq.~(9){]} agrees with the data using
the same $\gamma$ and $\mathcal{C}$ and no other fit parameters.
\textbf{c-d}. Optical excitation of noble-gas spin coherence.  \textbf{c}. On resonance,
the light field excites the noble-gas spins, which steadily precess with amplitude $\mathrm{R}_{0x}$  normalized to unity at maximal value [see Eq.~(11)]. A Lorentzian profile with linewidth $2\gamma$ (solid line) fits the data well. \textbf{d.} Measured precession signals on and off resonance. In \textbf{a-b}, each
data point is obtained from fitting a single heterodyne measurement
to a shifted harmonic signal (bars represent the fit confidence interval
of $95\%$). Error bars in \textbf{c} are below the marker size.}
\end{figure*}

The resonant interaction of the noble-gas spins with light also sets
the spins in motion and subjects them to coherent precession transversely
to the magnetic field. To demonstrate this optical excitation, we
send a long signal pulse (duration $T=3/\gamma$) and subsequently, while
keeping the control field on, monitor the precession of the noble-gas
spins using the polarization rotation of the control
field. 
The readout $R_\perp(t)$, proportional predominantly to the $x$-projection of the noble-gas spin $\bf{R}$, oscillates at the frequency $\omega_\mathrm{b}$, as shown in Fig.~2d (see Methods). We denote by $R_{0\perp}$ the amplitude of these oscillations and present $|R_{0\perp}|^2$ as a function of the optical signal frequency in Fig.~2c. As expected, the response of the spins to the optical excitation follows a narrow Lorentzian profile of linewidth $2\gamma$ .

The sharp variation of the optical susceptibility and the excitation of spin coherence near the resonance of the noble-gas spins are evidence for the emergent interaction between noble-gas spins and light. We now turn to model this interaction and study its dependence on the
applied magnetic field. The polarized alkali atoms, mediating this interaction, couple to the optical signal via the so-called Faraday Hamiltonian \cite{Polzik-RMP-2010} and couple magnetically to the noble-gas spins. The magnetic coupling is dominated by frequent spin-exchange collisions, whose accumulative effect leads to asymmetric shifts of the Larmor frequencies and, importantly, to coherent exchange between the transverse components of the two spins (in the $xy$ plane) at a rate $J$ \cite{Firstenberg-Weak-collisions}. See Methods for
the detailed model.

The evolution under the joint optical and exchange interactions can
be described in the following picture. The oscillating quadrature $S_{3}$ of the light field (pertaining to circular polarization) drives the transverse
alkali spin component $F_{y}$ at the frequency $\omega$, via a light-induced
shift. The transverse alkali spin components ($F_{x}$,$F_{y}$) couple to the
transverse noble-gas spin components ($R_{x}$,$R_{y}$) by the exchange interaction.
Therefore, the noble-gas spin can be resonantly excited when $\omega$ is tuned near its resonance frequency $\omega_{\mathrm{b}}$. Since the process is detuned from the alkali resonance frequency $\omega_{\mathrm{a}}$, the alkali spin effectively hybridizes with the noble-gas spin and adiabatically follows the noble-gas spin precession \cite{relax2}. As a back action, the precessing alkali spin -- specifically its oscillating component $F_{x}$ along the
optical axis -- rotates the light polarization and therefore changes the amplitude and phase of $S_2(\omega)$.

Using this picture, we theoretically calculate in Methods the output light quadrature
$S_{2}^{\mathrm{out}}(\omega)$. When the alkali spin
resonance is far detuned, \emph{i.e.}, when $|\delta_{\mathrm{a}}|\gg \gamma_{\mathrm{a}}$, where  $\delta_{\mathrm{a}}\equiv\omega-\omega_{\mathrm{a}}$,
we find that $S_{2}^{\mathrm{out}}(\omega)$ around the noble-gas
resonance is an inverted complex Lorentzian with a half width
\begin{equation}\gamma=\gamma_{\mathrm{b}}+\frac{J^{2}}{\delta_{\mathrm{a}}^{2}+\gamma_{\mathrm{a}}^{2}}\gamma_{\mathrm{a}}.\label{eq:gamma_definition}
\end{equation}
The first term $\gamma_{\mathrm{b}}$ is the bare decoherence rate of
the noble-gas spins. The second term is due to the hybridization of
the noble-gas and alkali spins, with the former inheriting a fraction
$J^{2}/(\delta_{\mathrm{a}}^{2}+\gamma_{\mathrm{a}}^{2})$ of the alkali
decoherence rate. This fraction is primarily set by the magnetic field
via $\delta_{\mathrm{a}}(B)$. We furthermore derive the contrast $\mathcal{C}$
of the spectral line (the maximal attenuation) and find that it, as well, decreases with increasing $|\delta_{\mathrm{a}}|$,
as the coupling mediated by the alkali spins weakens [see~Eqs.~(8)
and (10)]. It is therefore expected that decreasing the
magnetic field will increase the contrast, broaden the spectral
line, and enhance the spin precession signal.

To verify the model, we measure optical spectra for varying magnetic
fields and present the extracted linewidth and contrast in Fig.~3.
We fit the data to our model and find an excellent
agreement with the dependence on the magnetic field, for $J=14\pm0.4~\mathrm{Hz}$
and $2\gamma_{\mathrm{b}}=4.8\pm1~\mathrm{mHz}$, the latter mainly
limited in our setup by nonuniformity of the equivalent magnetic field of the gases over the cell volume. The alkali parameters $\gamma_{\mathrm{a}}=51\pm3~\textrm{Hz}$ and $\delta_{\mathrm{a}}(B)$
are calibrated independently (see Methods). We obtain the narrowest line for $B=10.7~\mathrm{mG}$, with a width of $2\gamma=5\pm0.7~\mathrm{mHz}$
that approaches the bare decoherence rate $2\gamma_{\mathrm{b}}$. We note that the basic constraints that connect $B$ to $\delta_{\mathrm{a}}$ and determine the degree of coupling could potentially be circumvented by modulating the magnetic field near the alkali spin resonance frequency \cite{VOLK}.

Finally, we use a magnetic pulse excitation to independently measure the transverse relaxation rate of the noble-gas
spins in the absence of a signal field. The data, presented in Fig.~3a
(diamonds), attest that the noble-gas decoherence rate is insensitive to the amplitude of the signal.

\begin{figure}[t]
\begin{centering}
\includegraphics[viewport=0bp 0bp 798bp 865bp,clip,width=6.5cm]{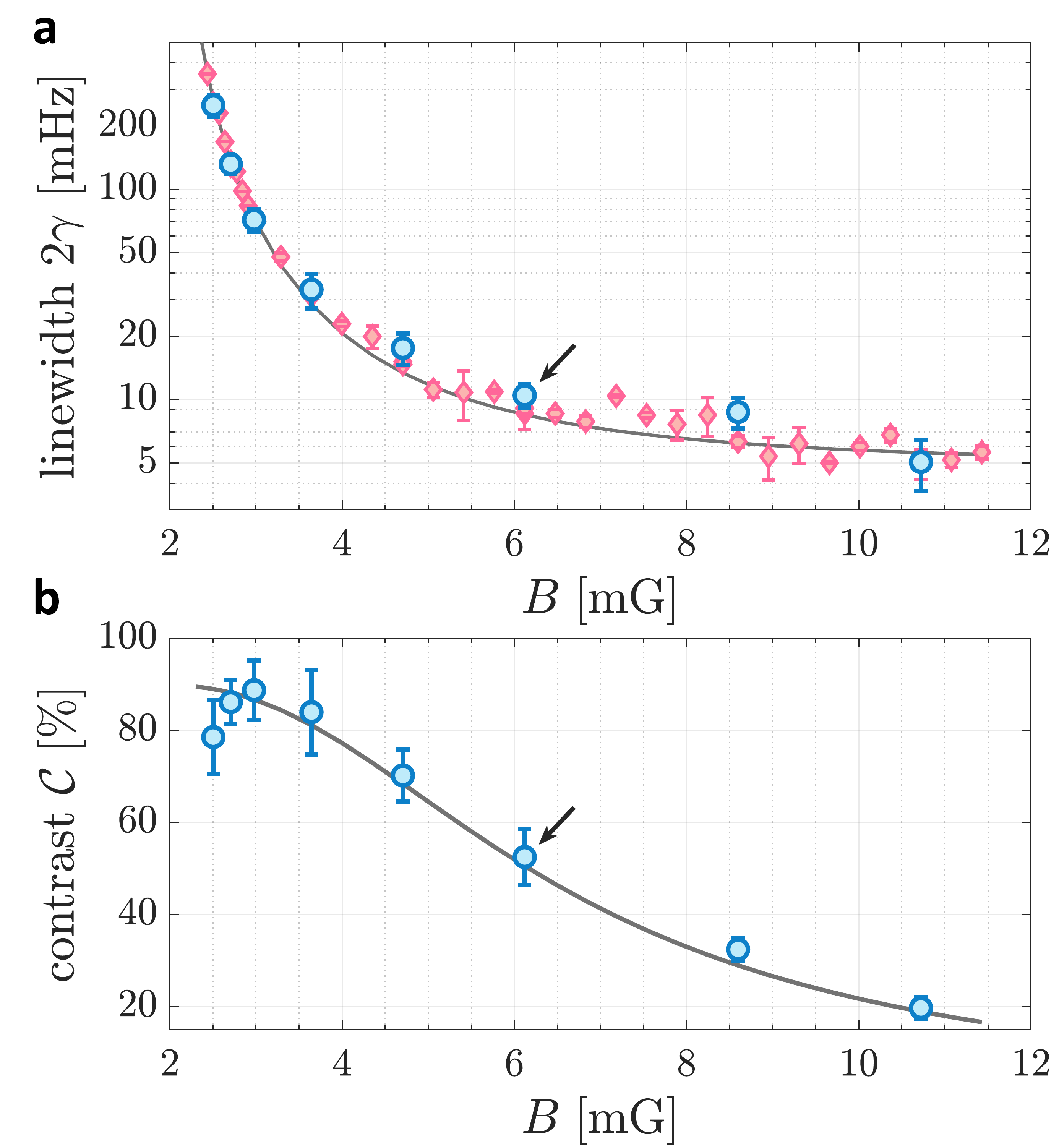}
\par\end{centering}
\centering{}\caption{\textbf{Magnetic control of the optical spectral line.} \textbf{(a)} Narrowing of the line and \textbf{(b)} decrease of the line contrast 
with increasing magnetic field. Circles are spectroscopic data and lines are the theoretical model, which describes the hybridization
of the noble-gas and alkali spins.
Stronger hybridization (low magnetic field) leads to stronger optical response at the noble-gas resonance on the expense of faster relaxation of the hybridized spin (dominated by the alkali relaxation). Black arrows indicate
the measurements presented in Fig.~2. In \textbf{(a)}, the spectroscopically measured linewidths agree with independent
transient measurements of the relaxation rate of the noble-gas spins (diamonds). The measured linewidth at $B=10.7~\mathrm{mG}$ is $2\gamma=5\pm0.7~\mathrm{mHz}$, approaching the bare noble-gas decoherence rate in our setup.
Error bars represent a confidence interval of $95\%$. \label{fig: magnetic_field_dependence}}
\end{figure}

\section*{Discussion}
It is insightful to compare the magneto-optical response we employ in this study with the response used in noble-gas NMR sensors, such as NMR gyroscopes \cite{Walker-NMRG,sensor}. While the former is linear in the signal field and nonlinear in the control field, \textit{i.e.}, a $\chi^{(3)}$ process (nonlinear susceptibility), the latter is linear in the control field only, which is a $\chi^{(1)}$ process (linear susceptibility) \cite{NMOR,nmor3,nmor2}. Therefore in NMR sensors, the rotation of the control field appears as an additive (source) term in the outgoing signal field, which is independent of the incoming signal amplitude. In contrast, the $\chi^{(3)}$ process employed in this work produces a homogeneous term in the propagation equations of the signal. This homogeneous term, corresponding to a linear coupling between the signal and the spins, is desired for quantum optics applications. It indicates that the signal field excites the spins, as oppose to NMR sensors, where the spins are excited by an external perturbation (namely rotation or magnetic field). High efficiency of the bi-directional interface requires a large bi-directional coupling rate $J=\sqrt{J_{\mathrm{a}}J_{\mathrm{b}}}$, defined in terms of the uni-directional coupling rates $J_{\mathrm{a}}$ and $J_{\mathrm{b}}$ [see Eq.~(3) in Methods], which corresponds to maximizing the spin-exchange fields of both gases. In contrast, the uni-directional operation of NMR sensors relies only on $J_{\mathrm{a}}$, which depends solely on the spin-exchange field of the noble gas.
A bi-directional optical interface to noble-gas spins can be used to couple these spins to distant objects, such as done in hybrid optomechanical-atomic-spin systems \cite{optoMech,optoMech2,optoMech3}. 

Quantum memories, in particular, require several extensions to our scheme, which are readily available.
First, quantum memories necessitate the coherent mapping of both quadratures of the signal field on the spins \cite{Polzik-RMP-2010}, whereas, in this work, only one quadrature couples coherently to the noble-gas spins. To realize a complete mapping in the presence of high noble-gas pressure, the so-called double-pass interaction can be employed, by sending the light beam twice through the cell \cite{Polzik2006,Cirac2006}. Second, quantum memories require a high bandwidth of the mapping process compared to the decoherence rate of the spins. A large time-bandwidth product usually requires temporal variations of external classical fields. Indeed, variations of the magnetic and control fields can be done during the reading and writing processes to increase the bandwidth and, during the memory time, to decouple the noble-gas spins from the alkali and consequently reduce its decoherence rate down to $\gamma_{\mathrm{b}}$ \cite{noble-storage-light,noble-storage-light2}.
    
In summary, we realize an efficient bi-directional coupling between noble-gas spins and light. The optical field coherently excites the long-lived noble-gas spins,
which in turn alter the field, generating
sharp complex-Lorentzian spectral lines. We model the coupling via hybridization of the noble-gas spins with the alkali spins and show that the magnetic field controls the degree of hybridization and thus the contrast and width of the optical lines. It is remarkable that the signal field is attenuated so efficiently by the mixture of alkali and noble-gas spins, despite each of them separately being transparent to light at the signal frequency. As this optical coupling exploits the long-lived coherence of the noble-gas spins, it could have various applications in precision measurements
of magnetic or anomalous fields \cite{Walker-RMP-2017,NP-1} and in the generation of long-lived entanglement and optical quantum memories
for quantum sensing and quantum information processing \cite{Firstenberg-QND,noble-storage-light}.

\section*{Acknowledgments:}
We thank Mark Dikopoltsev and Or Tal for their assistance in constructing the magnetic setup and setting the remote control, and Chen Avinadav, Ran Finkelstein, and Or Peleg for fruitful discussions. \textbf{Funding:} We acknowledge financial support by the Israel Science Foundation, the European Research Council starting investigator grant Q-PHOTONICS 678674, the Pazy Foundation, the Minerva Foundation with funding from the Federal German Ministry for Education and Research, and the Laboratory in Memory of Leon and Blacky Broder.

\bibliography{scibib}

\bibliographystyle{Science}

\clearpage

\section*{Materials and Methods}

\subsection*{Detailed Experimental Configuration}

We use an aluminosillicate spherical glass cell
of diameter $14$ mm containing 1500 Torr of
$^{3}\text{He}$ atoms and a droplet of potassium metal. The cell
also contains 40 Torr of $\textrm{N}_{2}$ gas for quenching
and mitigation of radiation trapping. We use twisted-pair resistance
wires with current oscillating at $450~\mathrm{kHz}$ to stabilize
the cell temperature to $T=187\,^{\circ}\textrm{C}$, yielding an estimated
potassium density of $n_{\mathrm{a}}\approx8.5\times10^{13}\,\textrm{cm}^{-3}$.
A constant magnetic field $B\hat{z}$ is generated using a set of
coils located within five layers of magnetic shields. The magnetic
shields are de-gaussed and the transverse magnetic field is zeroed
with an additional set of coils. The optical-pumping light originates
from a single-mode, linearly-polarized, distributed Bragg reflector
(DBR) laser at $770~\mathrm{nm}$. It is amplified with a tapered amplifier
and stabilized to $500\,\textrm{mW}$ using a commercial noise eater and passes through a $\lambda/4$ retarder to render its polarization circular.
The pumping light is blue-detuned from the $D_{1}$
optical line by $150~\mathrm{GHz}$ and fills the entire cell (estimated
beam waist diameter of $10~\textrm{mm}$); these measures increase the spatial
homogeneity of the alkali spin polarization.

We polarize the helium spins along $\hat{z}$ at a magnetic field
of $B=37\,\textrm{mG}$ using spin-exchange optical pumping (SEOP).
The equivalent magnetic field (EMF) exerted by the helium on the potassium spins is built at a typical rate of $1~\mathrm{mG/hour}$,
as measured by using the potassium as a magnetometer. The longitudinal
lifetime of the helium spins is greater than 2.5 hours, plausibly
limited by nonzero transverse magnetic field exerted by polarized
helium atoms residing in the stem of the cell. The helium polarization is kept constant by a slow servo loop, which monitors
the helium EMF, controls the pumping rate, and maintains the helium EMF at $2.8~\textrm{mG}$
to better than one percent. The decoherence rate of the helium spins
$2\gamma_{\mathrm{b}}=4.8\pm1~\textrm{mHz}$ is predominantly limited
by spatial nonuniformity of the EMF of both species. To
initialize each measurement with zero coherence of the collective
helium spin, we apply a smooth pulse of magnetic-field gradient along
$\hat{z}.$ This dephases any precession of helium spin remnant from
prior measurements due to its long coherence time.

The signal and control fields originate from another single-mode DBR
laser at $770~\mathrm{nm}$. They are split using a polarizing beam splitter (PBS), sent to independent
acousto-optic modulators, and the modulated beams are recombined using
a second PBS. In the combined beam, the signal is typically $50~\textrm{dB}$
weaker than the control, and we vary the frequency difference between
them $(\omega)$ through the relative RF frequencies of the two modulators.
We use a retarder and a precision Soleil-Babinet compensator to align
the linear polarization of the control field with the magnetic field.
The input Gaussian beam, with $25~\textrm{mW}$, is expanded to a waist
of $7~\textrm{mm}$ and slightly focused to compensate for lensing
by the spherical cell. 

We measure the $S_{2}$ quadrature of the output beam $S_{2}^{\mathrm{out}}(t)$ using a polarization
heterodyne detection setup, comprising a zero-order $\lambda$/2 retarder
at ${\sim}22.5^{\circ}$, a PBS, and a differential photodetector. The
detector outputs are subtracted and amplified, producing a readout
proportional to $S_{2}^{\mathrm{out}}(t)$. The exact angle of the
retarder is set to null the readout in the presence of only the control
field, {\it i.e.}, to balance the detection. Digital Fourier transform is used to obtain $S_{2}^{\mathrm{out}}(\omega)$.
For normalization and for extraction of the phase shift $\varphi(\omega)=\arg[S_{2}^{\mathrm{out}}(\omega)/S_{2}^{\mathrm{in}}(\omega)]$,
we also sample the input beam before the cell and measure $S_{2}^{\mathrm{in}}(\omega)$
using a similar setup. To measure the optically-excited nuclear spin precession after turning off the signal, we employ the alkali spins as an effective magnetometer, probed by polarization rotation of the control field. The polarization rotation is linearly proportional to $F_x(t)$, which in turn is proportional to the  noble-gas spin $R_\perp(t)$ oscillaing in the $xy$ plane, as inferred from Eq.~(\ref{eq:SteadyF}). This harmonic signal has  an amplitude $R_{0\perp}(\Delta)$ which is plotted in Fig.~2c. The scale of $R_{0\perp}(\Delta)$ is normalized such that $R_{0\perp}(\Delta=0)=1$.

\subsection*{Fitting and Calibrations}

Frequency variables throughout the paper indicate angular frequencies, and $\mathrm{Hz}=2\pi\,\mathrm{rad}/\mathrm{s}$.
We independently calibrate $\gamma_{\mathrm{a}}$ and $\omega_{\mathrm{a}}(B)$
by measuring the precession of the potassium spins. We slightly tilt
the spins by applying a short pulse of magnetic field along $\hat{y}$
and monitor their precession using polarization rotation measurements in
the presence of the pumping field and the magnetic field $B\hat{z}$.
A fit of the precession frequency to the linear function $\omega_{\mathrm{a}}(B)=g_{\mathrm{a}}(B-B_{0}^{\mathrm{b}})$
yields an estimate of the helium EMF  $B_{0}^{\mathrm{b}}$
and the effective gyromagnetic ratio $g_{\mathrm{a}}=592\,\textrm{Hz/mG}$
of the potassium. The latter is slower than the gyromagnetic ratio
of a free electron by a factor of $q_{\mathrm{a}}=4.7$, corresponding
to an estimated potassium polarization of $p_{\mathrm{a}}=70\%$ in
the regime of rapid potassium-potassium collisions (see Ref.~\cite{happer-1998};
theoretically, $q_{\mathrm{a}}=2I+1=4$ in the absence of such collisions
or when $p_{\mathrm{a}}=100\%$, and $q_{\mathrm{a}}=6$ for $p_{\mathrm{a}}=0$).
From these precession measurements, we also infer the decoherence rate
of the potassium spins at low magnetic fields $\gamma_{\mathrm{a}}=51\pm3~\textrm{Hz}$
in the presence of both pumping and control light. 

\subsection*{Theoretical Model}

The complex amplitude of the electric field of the combined control
and signal field is given by $\mathcal{\boldsymbol{E}}=\mathcal{E}_{\mathrm{c}}\hat{z}+(e^{i\omega t}\mathcal{E}_{+}+e^{-i\omega t}\mathcal{E}_{-})\hat{y}$.
Here $\mathcal{E}_{\mathrm{c}}$ is the amplitude of the control field,
and $\mathcal{E}_{\mathrm{+}}$ is the amplitude of the signal field
at the entrance to the cell, where $\mathcal{E}_{-}=0$. These amplitudes vary along the cell. We describe the signal field before and after the cell in terms of the Stokes parameters $S_{2}(t)+iS_{3}(t)=2(\mathcal{\boldsymbol{E}}\cdot\hat{z})^{*}(\mathcal{\boldsymbol{E}}\cdot\hat{y})$
and their spectral representation $S_{2}(\omega)=\mathcal{E}_{\mathrm{c}}^{*}\mathcal{E}_{-}+\mathcal{E}_{\mathrm{c}}\mathcal{E}_{+}^{*}$
and $S_{3}(\omega)=i(\mathcal{E}_{\mathrm{c}}\mathcal{E}_{+}^{*}-\mathcal{E}_{\mathrm{c}}^{*}\mathcal{E}_{-})$.

\subsubsection*{Effect of light on atoms}

The equations of motion for the transverse mean spin components of
the alkali and noble-gas atoms comprise the Faraday and spin-exchange
Hamiltonians \cite{Firstenberg-QND} 
\begin{align}
\partial_{t}\boldsymbol{\mathrm{F}} & =\bigl(\omega_{\mathrm{a}}\boldsymbol{\mathrm{F}}-J_{\mathrm{a}}\boldsymbol{\mathrm{R}}\bigr)\times\hat{z}-\gamma_{\mathrm{a}}\boldsymbol{\mathrm{F}}+\bar{a}p_{\mathrm{a}}S_{3}(t)\hat{y},\\
\partial_{t}\boldsymbol{\mathrm{R}} & =\bigl(\omega_{\mathrm{b}}\boldsymbol{\mathrm{R}}-J_{\mathrm{b}}\boldsymbol{\mathrm{F}}\bigr)\times\hat{z}-\gamma_{\mathrm{b}}\boldsymbol{\mathrm{R}}.\label{eq:noble-gas_equation}
\end{align}
Here $\boldsymbol{\mathrm{F}}$ and $\boldsymbol{\mathrm{R}}$ denote,
respectively, the mean alkali and noble-gas spin vectors in the $xy$
plane, which experience Larmor precession at frequencies $\omega_{\mathrm{a}}$
and $\omega_{\mathrm{b}}$ and are coupled by the uni-directional coherent spin-exchange
rates $J_{\mathrm{a}}=q_{\mathrm{a}}\zeta n_{\mathrm{b}}p_{\mathrm{a}}/2$ and $J_{\mathrm{b}}=\zeta n_{\mathrm{a}}p_{\mathrm{b}}/2$.
These rates depend on the number densities $n_{\mathrm{a}},\,n_{\mathrm{b}}$
and on the degree of polarization $0\leq p_{\mathrm{a}},\,p_{\mathrm{b}}\leq1$
of the alkali and noble-gas spins, respectively, and on $\zeta(q_{\mathrm{a}}=4.7)=4\times10^{-15}\,\textrm{cm}^{3}\textrm{s}^{-1}$
\cite{Firstenberg-Weak-collisions}. The symmetric (bi-directional)
coupling rate in the bosonic representation is $J=\sqrt{J_{\mathrm{a}}J_{\mathrm{b}}}$
\cite{Firstenberg-Weak-collisions}.
The circularly polarized component $S_{3}(t)$ of the combined control
and signal field exerts light-shift on the alkali spin. In the far-detuned
limit $|\delta_{\mathrm{e}}|\gg\gamma_{e}$, the light shift tilts the
spin (predominantly its $\hat{z}$ component) around
the $\hat{x}$ axis at a rate $\bar{a}=2r_{e}c/(3q_{\mathrm{a}}A\delta_{\mathrm{e}})$.
Here $r_{e}=2.8\times10^{-13}\,\textrm{cm}$ is the electron radius,
$c$ is the speed of light, and $A$ is the beam area.

Transforming the alkali and noble-gas spin vectors to a complex coherence
representation in the rotating frame $\tilde{F}(t)=\boldsymbol{F}(t)(\hat{x}+i\hat{y})e^{i\omega t}$
and $\tilde{R}(t)=\boldsymbol{\mathrm{R}}(t)(\hat{x}+i\hat{y})e^{i\omega t}$,
we carry out the rotating-wave approximation and find, in the steady
state $\partial_{t} \tilde{F}=\partial_{t}\tilde{R}=0$, 
\begin{equation}
\tilde{F}=\frac{i\bar{a}p_{\mathrm{a}}S_{3}(\omega)}{\gamma_{\mathrm{a}}-i\delta_{\mathrm{a}}+\frac{J^{2}}{\gamma_{\mathrm{b}}-i\delta_{\mathrm{b}}}}\label{eq:alkali_coherence}
\end{equation}
and
\begin{equation}
\tilde{R}=\frac{-iJ_{\mathrm{b}}\bar{a}p_{\mathrm{a}}S_{3}(\omega)}{\gamma_{\mathrm{a}}-i\delta_{\mathrm{a}}}\frac{1}{\gamma-i\Delta}.\label{eq:K_coherence}
\end{equation}
Equation (\ref{eq:alkali_coherence}) describes the alkali coherence
driven by the signal field ($S_{3}$ component). The coupling to the
noble-gas is manifested as a complex Lorentzian $J^{2}/(\gamma_{\mathrm{b}}-i\delta_{\mathrm{b}})$
added to the denominator. Importantly, for slowly varying frequencies $\omega\ll\omega_{\mathrm{a}}$,
the rate $\gamma_{\mathrm{a}}$ is free from spin-exchange relaxation
due to alkali-alkali collisions \cite{happer-1998}. Here the alkali
spins adiabatically follow the spin precession of the noble-gas spins,
constantly oriented along the effective magnetic field they
experience; spin-exchange relaxation affects only the spin components
transverse to this effective magnetic field. 

In the limit $|\delta_{\mathrm{a}}|\gg\gamma_{\mathrm{a}}$ realized in
the experiment, this additional complex-Lorentzian generates a Lorentzian
Raman resonance with a narrow linewidth $2\gamma$. The noble-gas
spin coherence in Eq.~(5) is excited by the
alkali-spin mediator and dominated as well by a complex Lorentizan function,
with the width $2\gamma$ {[}Eq.~(1){]}
and detuning $\Delta=\delta_{\mathrm{b}}-J^{2}\delta_{\mathrm{a}}/(\delta_{\mathrm{a}}^{2}+\gamma_{\mathrm{a}}^{2})$,
as indeed we measure for $|\delta_{\mathrm{a}}|\gg\gamma_{\mathrm{a}}$ (Fig.~2c).

\subsubsection*{Effect of atoms on light}

The combined control and signal beam traversing
the atomic medium experiences polarization rotation 
\begin{equation}
S_{2}^{\mathrm{out}}(\omega)=S_{2}^{\mathrm{in}}(\omega)+\alpha\tilde{F},\label{eq:faraday rotation}
\end{equation}
while $S_{3}^{\mathrm{out}}(\omega)=S_{3}^{\mathrm{in}}(\omega)$ remains
constant. Here $\alpha=dn_{\mathrm{a}}\bar{a}AP_{\mathrm{c}}/(4\hbar\omega_{\mathrm{e}})$,
where $P_{\mathrm{c}}$ is the power of the control field, and $\hbar\omega_{\mathrm{e}}$
is the photon energy. Substituting the alkali coherence Eq.~(\ref{eq:alkali_coherence})
in the limit $|\delta_{\mathrm{a}}|\gg\gamma_{\mathrm{a}}$ into Eq.~(\ref{eq:faraday rotation})
yields the optical spectral response 
\begin{equation}
S_{2}^{\mathrm{out}}(\omega)=S_{2}^{\mathrm{in}}(\omega)\left(1-\mathcal{C}_{0}\frac{\gamma}{\gamma-i\Delta}\right),\label{eq:S2_spectrum}
\end{equation}
where 
\begin{equation}
\mathcal{C}_{0}=\frac{p_{\mathrm{a}}\textrm{OD}}{2}\frac{\gamma'_{\mathrm{a}}}{\gamma_{\mathrm{a}}}\frac{\gamma-\gamma_{\mathrm{b}}}{\gamma}\label{eq:C0}
\end{equation}
is the amplitude of the complex Lorentzian. It depends on the hot,
on-resonance optical depth ($\textrm{OD}$) and on the relaxation of the alkali
spins due to scattering of control photons $\gamma'_{\mathrm{a}}=\alpha\bar{a}/\textrm{OD}$,
with respect to the total relaxation rate $\gamma_{\textrm{a}}$.
From Eq.~(\ref{eq:S2_spectrum}), we obtain the phase shift
\begin{equation}
\varphi(\omega)\equiv\textrm{arg}\frac{S_{2}^{\mathrm{out}}(\omega)}{S_{2}^{\mathrm{in}}(\omega)}=\textrm{atan}\left(\frac{\mathcal{C}_{0}\gamma\Delta}{\Delta^{2}+\gamma^{2}(1-\mathcal{C}_{0})}\right),\label{eq:phase plot}
\end{equation}
and we find that $|S_{2}^{\mathrm{out}}(\omega)/S_{2}^{\mathrm{in}}(\omega)|^{2}$
is exactly an inverted Lorentzian, with width $2\gamma$ and contrast
\begin{equation}
\mathcal{C}=\mathcal{C}_{0}(2-\mathcal{C}_{0}),\label{eq:C}
\end{equation} as presented in Figs.~2(a,b).

For the experiments presented in Figs.~2c and 2d, the input signal field is turned off and the control field is kept on to act as a monitor. The noble-gas spins precess at their bare frequency $\omega_\mathrm{b}$, and the alkali spins adiabatically follow. The $x$-projection of the alkali spins, in a frame oscillating at a frequency $\omega_\mathrm{b}$, determines the rotation of the control field and is given by
\begin{equation}
F_x(t)=\frac{J_\mathrm{a}}{\sqrt{(\omega_{\mathrm{a}}-\omega_{\mathrm{b}})^2+\gamma_{\mathrm{a}}^2}}R_\perp(t),\label{eq:SteadyF}
\end{equation}
where  $R_\perp(t)=\mathrm{Re}\left[\boldsymbol{\mathrm{R}}(t)(\cos\psi\hat{x}+i\sin\psi\hat{y})e^{i\omega_\mathrm{b} t}\right]$ is the noble-gas projection which depends on the parameter $\sin\psi=\gamma_{\mathrm{a}}/\sqrt{(\omega_{\mathrm{a}}-\omega_{\mathrm{b}})^2+\gamma_{\mathrm{a}}^2}$ . At our moderate magnetic fields, in which the alkali is far detuned $|\omega_{\mathrm{a}}-\omega_{\mathrm{b}}|\gg\gamma_{\mathrm{a}}$ (so that $\sin\psi\ll\cos\psi$), the measurement is predominantly sensitive to the $x$ component of the noble-gas spins.

\end{document}